\documentclass{article}
\usepackage{amsfonts, amsmath, amsthm, amssymb,spconf,graphicx}
\usepackage{booktabs}

\usepackage[normalem]{ulem}
\usepackage{soul}

\title{Automatic Extraction and Sign Determination of Respiratory Signal in Real-time Cardiac Magnetic Resonance imaging}
\name{Chong Chen, Yingmin Liu, Orlando P. Simonetti, Rizwan Ahmad\thanks{Corresponding author: Rizwan Ahmad (ahmad.46@osu.edu). This work was funded by NIH R01HL135489.}}
\address{The Ohio State University}
\usepackage{color}
\newcommand{\textr}[1]{\textcolor{black}{#1}}

\newcommand{\Real}{{\mathbb{R}}}
\newcommand{\argminE}{\mathop{\mathrm{argmin}}}          

\begin{document}
%
\maketitle
\begin{abstract}
In real-time (RT) cardiac cine imaging, a stack of 2D slices is collected sequentially under free-breathing conditions. A complete heartbeat from each slice is then used for cardiac function quantification. The inter-slice respiratory mismatch can compromise accurate quantification of cardiac function. Methods based on principal components analysis (PCA) have been proposed to extract the respiratory signal from RT cardiac cine, but these methods cannot resolve the inter-slice sign ambiguity of the respiratory signal. In this work, we propose a fully automatic sign correction procedure based on the similarity of neighboring slices and correlation to the center-of-mass curve. The proposed method is evaluated in eleven volunteers, with ten slices per volunteer. The motion in a manually selected region-of-interest (ROI) is used as a reference. The results show that the extracted respiratory signal has a high, positive correlation with the reference in all cases. The qualitative assessment of images also shows that the proposed approach can accurately identify heartbeats, one from each slice, belonging to the same respiratory phase. This approach can improve cardiac function quantification for RT cine without manual intervention. 

\end{abstract}
\section{Introduction}
\label{sec:intro}
Respiratory motion is one of the main sources of artifacts and inaccurate quantification in cardiac imaging. To reduce its influence, breath-held or respiratory-gated protocols are widely used. However, these techniques have limitations for patients with arrhythmia or poor respiratory control. Moreover, these techniques are unable to capture the respiratory motion-induced changes in the cardiac output. As a result, there is a growing demand to develop real-time (RT) cardiovascular magnetic resonance imaging (CMR) methods. 
Cardiac cine, i.e., creating a movie of the beating heart, is one of the most common applications of CMR. In RT cardiac cine, a stack of 2D slices is collected sequentially under free-breathing conditions. A complete heartbeat from each slice is then used for cardiac function quantification. 
To get accurate and reproducible cardiac output~\cite{claessen2014interaction}, the determination of the respiratory phase is essential. The respiratory signal can be obtained using an external device, such as respiratory bellows. However, they often need prior setup and do not always work reliably in patients. 


The motion information is already present in the RT cine images; therefore, the respiratory signal can be extracted directly using data-driven methods. Various strategies to extract the respiratory signal from RT cine data have been developed~\cite{cai2015extracting,thielemans2011device,novillo2019unsupervised}. Among them, PCA-based methods~\cite{thielemans2011device,novillo2019unsupervised} are particularly appealing because they are fast and independent of the slice orientation. However, due to the nature of PCA, the extracted signal has an arbitrary sign, which means that the 
maximal signal could refer either to the inspiration or expiration phase. Several methods to resolve this issue have been proposed. Bertolli et al. assumed the directionality of craniocaudal motion of the internal organs to resolve the sign ambiguity in PET images~\cite{bertolli2017sign}. In MRI, Felipe et al. developed a sign correction method by tracking the diaphragmatic motion or estimating the lung volume, but this method requires selecting the lung-diaphragm interface~\cite{novillo2019unsupervised}.

In this work, we propose and evaluate a fully automatic sign determination method that is based on the image content similarity of neighboring slices and correlation to the center-of-mass curve. The method is validated using data from 11 volunteers by both quantitative and qualitative measures. 

\begin{figure*}[htb]
  \centering
  \centerline{\includegraphics[width=17cm]{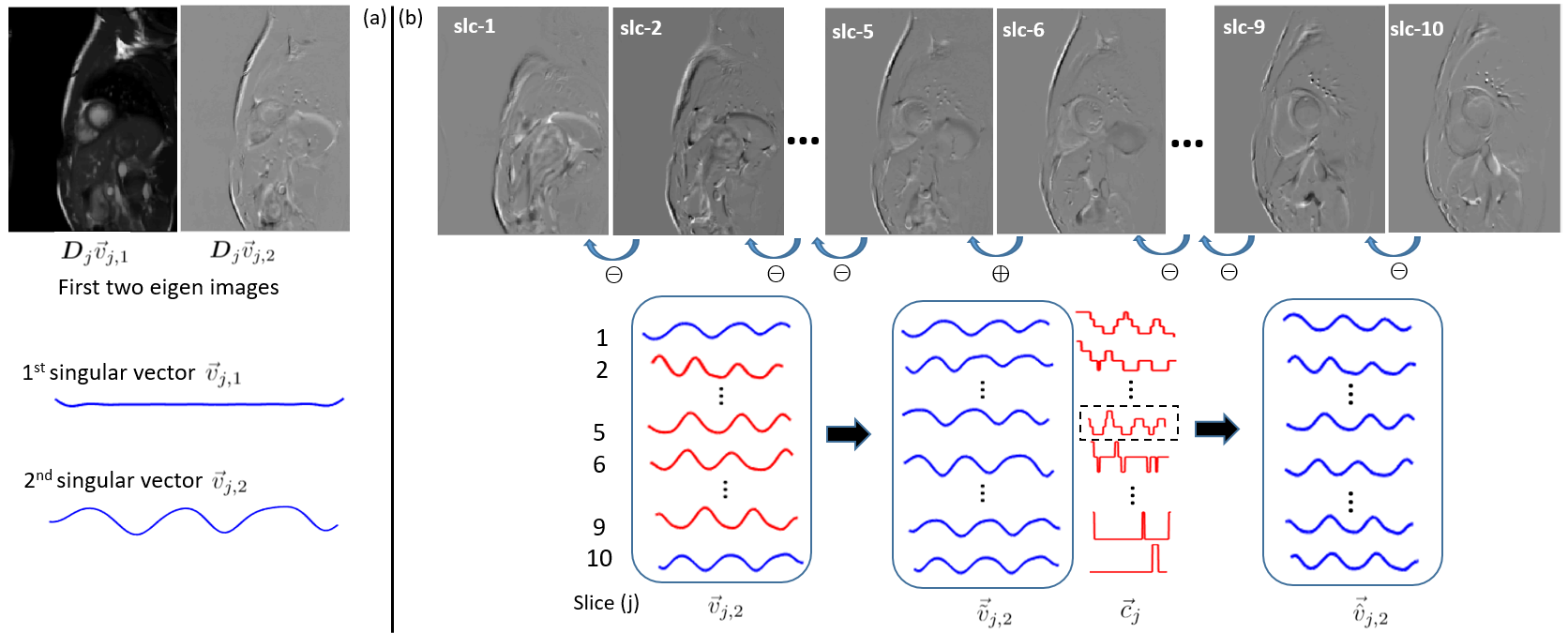}}
\caption{(a) The first two eigen images and right singular vectors. (b) Proposed sign determination procedure. First, a pair-wise inter-slice correlation is used to fix the sign of individual slices with respect to the first slice. Second, an overall sign correction is performed by maximizing the correlation of slices with the center-of-mass curve ($\vec{c}_j$). Here, slc represents slice.}
\label{fig:method}
\end{figure*}

\section{Methods}
\label{sec:format}
\subsection{Extraction of the respiratory signal}
Consider an image series with $n$ frames and $M$ pixels per frame. The image series from the $j^\text{th}$ slice can be represented as a matrix $\boldsymbol{D}_j\in\Real^{M\times n}$, with each column corresponding to a single frame. First, a low pass filter $[0-0.8\,\text{Hz}]$ is applied along the temporal dimension to suppress the higher frequency components, e.g., from the cardiac motion. Second, the eigen decomposition of $\boldsymbol{D}_j^\text{T} \boldsymbol{D}_j$ is performed to obtain right singular vectors, $\{ \Vec{v}_{j,1}, \vec{v}_{j,2},...,\vec{v}_{j,n} \}$, of $\boldsymbol{D}_j$. These singular vectors capture different motions present in $\boldsymbol{D}_j$, including the respiratory motion. Third,  $i^\text{th}$ ``eigen image'' from the $j^\text{th}$ slice is obtained by $\boldsymbol{D}_j\vec{v}_{j,i}$ and can be interpreted as a projection of the image series on the corresponding motion defined by $\vec{v}_{j,i}$. As shown in Fig.~\ref{fig:method} (a), the first singular vector, $\vec{v}_{j,1}$, corresponds to the temporal average and the second singular vector, $\vec{v}_{j,2}$, captures the respiratory motion. 
For further processing, we only consider the second eigen image, $\boldsymbol{D}_j\vec{v}_{j,2}$, and the second singular vector, $\vec{v}_{j,2}$, from each slice.

\subsection{Sign determination}
We adopt a two-step procedure to resolve sign ambiguity in the respiratory motion from multi-slice CMR. In the first step, we adjust the sign of all slices ($j=2,3,\dots,J$) with respect to the first slice ($j=1$). To this end, we estimate Pearson correlation coefficient, $r_{j,j+1}$, between $\boldsymbol{D}_j\vec{v}_{j,2}$ and $\boldsymbol{D}_{j+1}\vec{v}_{{j+1},2}$ and adjust the sign of $\vec{v}_{{j+1},2}$ based on $\vec{\tilde{v}}_{j+1,2}=\vec{v}_{j+1,2}\prod_{l=1}^{l=j} \text{sign}(r_{l,l+1})$, where $\vec{\tilde{v}}_{j+1,2}$ represents the singular vector from the $(j+1)^\text{th}$ slice after the sign correction. This process is repeated for all values of $j$ as depicted in Fig.~\ref{fig:method} (b). Upon the conclusion of this step, the sign of $\vec{\tilde{v}}_{j,2}$ is consistent across slices but is still arbitrary with respect to the directionality of the respiratory motion and thus cannot distinguish inspiration from expiration. In the second step, we use the center-of-mass curves, $\vec{c}_j \in\Real^{n\times 1}$, to perform the overall sign correction. Pearson correlation coefficient, $s_{j}$, between $\vec{\tilde{v}}_{j,2}$ and $\vec{c}_j$ is estimated for all values of $j$. Then, the sign of $\vec{\tilde{v}}_{j,2}$ is adjusted by $\vec{\hat{v}}_{j,2}=\text{sign}(s_l)\vec{\tilde{v}}_{j,2}$, where $\vec{\hat{v}}_{j,2}$ is the singular vector from the $j^\text{th}$ slice after the overall sign correction, and $s_l$ is the correlation coefficient with the maximal absolute value, i.e., $|s_{l}|=\text{max}\{|s_1|, |s_2|,\dots, |s_{J}|\}$. The $\vec{c}_j$ curve for the $j^{\text{th}}$ slice is computed by tracking the center-of-mass in the superior-inferior (SI) direction across $n$ frames, \textr{i.e., $\vec{c}_{j}(i) = \argminE_m | \sum_{l=1}^{m}\boldsymbol{P}_{j}(l,i) - \sum_{l=1}^{L}\boldsymbol{P}_{j}(l,i)/2 |$, where $i \in \{1,2,\dots,n \}$ represents the $i^{\text{th}}$ frame, and $\boldsymbol{P}_j \in\Real^{L\times n}$ is the SI projection of the image series from the $j^{\text{th}}$ slice, and $L$ is the number of pixels along the SI direction.} The movement of the thoracic and abdominal organs as a function of respiration is predictable in the SI direction~\cite{wang1995respiratory}. Considering that the orientation of the imaging plane is precisely known, the SI motion in the patient coordinate system can be expressed in terms of image coordinates. Therefore, $\vec{c}_j$ calculated from the image series can be used as a surrogate for the respiratory motion. Note, the quality of $\vec{c}_j$ as well as correlation of $\vec{c}_j$ with $\vec{\tilde{v}}_{j,2}$ can be poor for certain slices; see Fig.~\ref{fig:method} (b). As a result, this step by itself cannot be reliably used to adjust the sign for individual slices.

\subsection{Evaluation}
\begin{figure}[htb]
  \centering
  \centerline{\includegraphics[width=8.5cm]{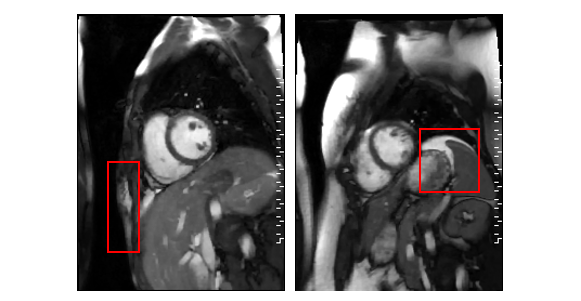}}
\caption{Manually placed ROI (red box) to extract the respiratory motion using non-rigid image registration. The images from two different volunteers are shown here. }
\label{fig:res}
\end{figure}

The motion in a manually selected region-of-interest (ROI) was used as a reference for the respiratory motion. As shown in Fig.~\ref{fig:res}, for each slice, the ROI was placed manually in an area \textr{with visually pronounced respiratory motion.} The frame-to-frame motion was captured using non-rigid image registration~\cite{chefd2002flows}. To this end, all frames for a given slice were registered to the first frame to produce 2D deformation fields with horizontal and vertical components. The deformation field characterized the motion of each pixel across $n$ frames. The respiratory signal was extracted by averaging the horizontal or vertical component of the deformation field within the ROI. The appropriate component of the deformation field was selected based on the visual assessment of the motion within the ROI. For example, for the ROI selected on the liver dome, the component of the deformation field in the head-foot direction would be chosen. Agreement between the respiratory signal from the proposed method, $\vec{\hat{v}}_{j,2}$, and the reference was evaluated using correlation coefficients. When the correlation was positive, the sign determination was considered to be correct.

\section{Experiment and results}
\label{sec:pagestyle}
\subsection{CMR scan setup}
Our method was evaluated in vivo using data from 11 healthy volunteers. A ten-slice short-axis stack of RT cine images covering the left ventricle was acquired for each volunteer on a 1.5\,T scanner (MAGNETOM  Avanto, Siemens Healthcare, Erlangen, Germany). \textr{Each slice was continuously acquired for 10 s to cover 1-3 respiratory cycles and 8-14 heartbeats. The slices were acquired sequentially, with a small time gap between slices. The starting respiratory phase for each slice was arbitrary and unknown.} The other imaging parameters were: \textr{slice thickness 8 mm, space between slices 10 mm}, FOV $300\times400$\,mm, TE/TR $1.18/2.83$\,ms, image matrix $192\times256$, flip angle $69^{\circ}$, acceleration rate $9$ using pseudo-random variable density sampling, spatial resolution $1.6\times1.6$\,mm$^2$, and temporal resolution $40$--$45$\,ms\textr{, resulting in 222--250 frames/slice.} The images were reconstructed inline using a compressed sensing method called SCoRe~\cite{chen2019sparsity}. \textr{For the recruitment and consent of human subjects used in this study, the ethical approval was given by an Internal Review Board (2005H0124) at The Ohio State University.}

\begin{table}[!htbp]
		\begin{center}
\begin{tabular}{ccc}
\toprule
\multicolumn{3}{c}{{\bf Pearson correlation coefficient}} \\
\midrule
    {Volunteer}     &     {Mean}   &  {Range} \\ \hline
   {\bf 1} &       0.96   &  0.91-0.98 \\

   {\bf 2} &       0.92   &  0.87-0.95 \\

   {\bf 3} &       0.94   &  0.86-0.98 \\

   {\bf 4} &       0.94   &  0.64-0.99 \\

   {\bf 5} &       0.92   &  0.82-0.99 \\

   {\bf 6} &       0.94   &  0.86-0.97 \\

   {\bf 7} &       0.93   &  0.86-0.98 \\

   {\bf 8} &       0.93   &  0.84-0.99 \\

   {\bf 9} &       0.92   &  0.73-0.98 \\

  {\bf 10} &       0.96   &  0.90-0.99 \\

  {\bf 11} &       0.97   &  0.88-0.99 \\
\bottomrule
\end{tabular}  
		\end{center}
	\caption{Pearson correlation coefficients between the extracted respiratory motion and the reference. Ten slices were scanned for each volunteer and the correlation coefficient was calculated slice-wise.}
	\label{tab:table_R}
\end{table}

\begin{figure}[htb]
  \centering
  \centerline{\includegraphics[width=8.4cm]{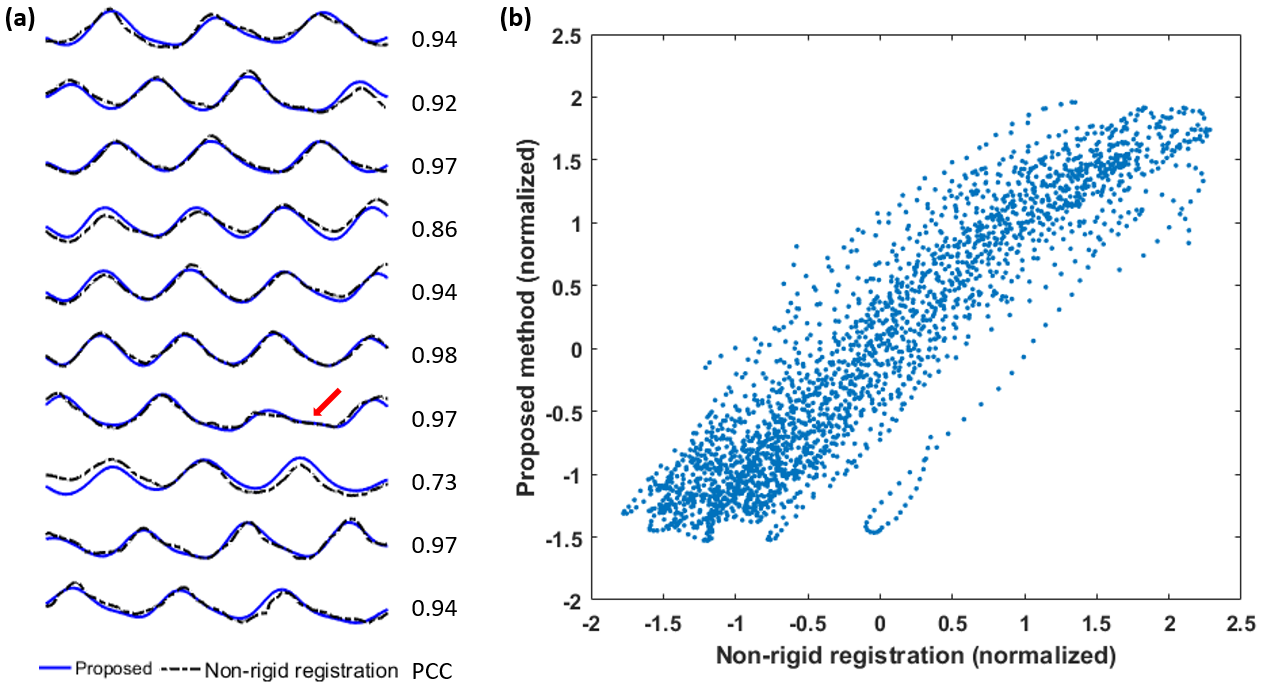}}
    \vspace{0.1cm}  
  \centerline{\includegraphics[width=8.4cm]{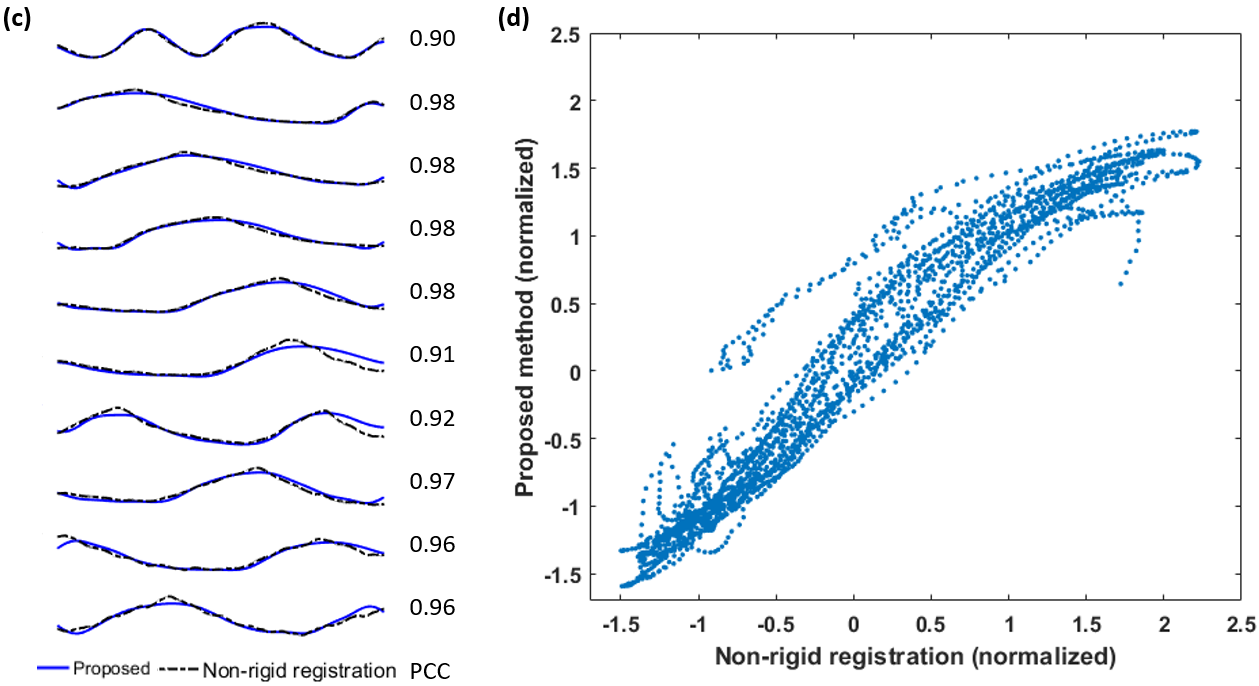}}
\caption{ The respiratory signal from the proposed method and the reference superimposed (a, c) as well as the corresponding scatter plots (b, d). \textr{PCC is the Pearson correlation coefficient between them. The results from two volunteers are shown, with the red arrow highlighting the irregular respiratory cycle}.}
\label{fig:res_curve}
\end{figure}

\subsection{Results}
Table \ref{tab:table_R} summarizes the Pearson correlation coefficients for all the volunteers. The results show that the extracted respiratory signal has a high, positive correlation with the reference in all cases. \textr{This method is also robust across different respiratory patterns}. Fig.~\ref{fig:res_curve} shows representative results from two volunteers, one with short respiratory cycles and the other with long cycles. \textr{As highlighted by the red arrow, this method is also tolerant of irregular motion}. For some slices in Fig.~\ref{fig:res_curve}, the difference between the reference and the proposed method is more pronounced. This is expected because the two methods measure different quantities: the proposed PCA-based method extracts the respiratory motion from the entire image, while the non-rigid image registration only tracks the motion in a small user-defined ROI. 

An important assumption in our method is the similarity between neighboring slices. To evaluate the impact of distance between adjacent slices, we effectively doubled the gap between adjacent slices by considering only odd slices. Even with the larger gap between neighboring slices, the proposed method yielded positive, albeit lower, correction with the reference in all cases.

\begin{figure}[htb]
  \centering
  \vspace{0.1cm}  
  \centerline{\includegraphics[width=8.4cm]{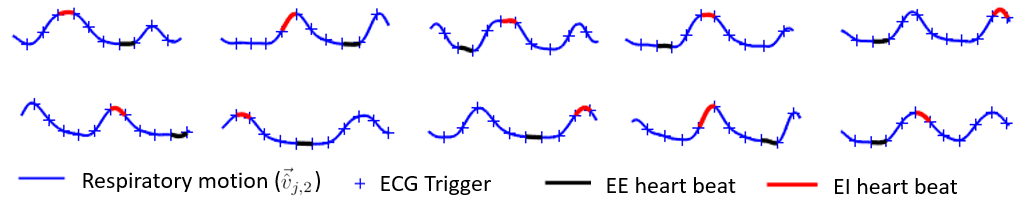}}  
  \vspace{0.0cm}
  \centerline{\includegraphics[width=8.4cm]{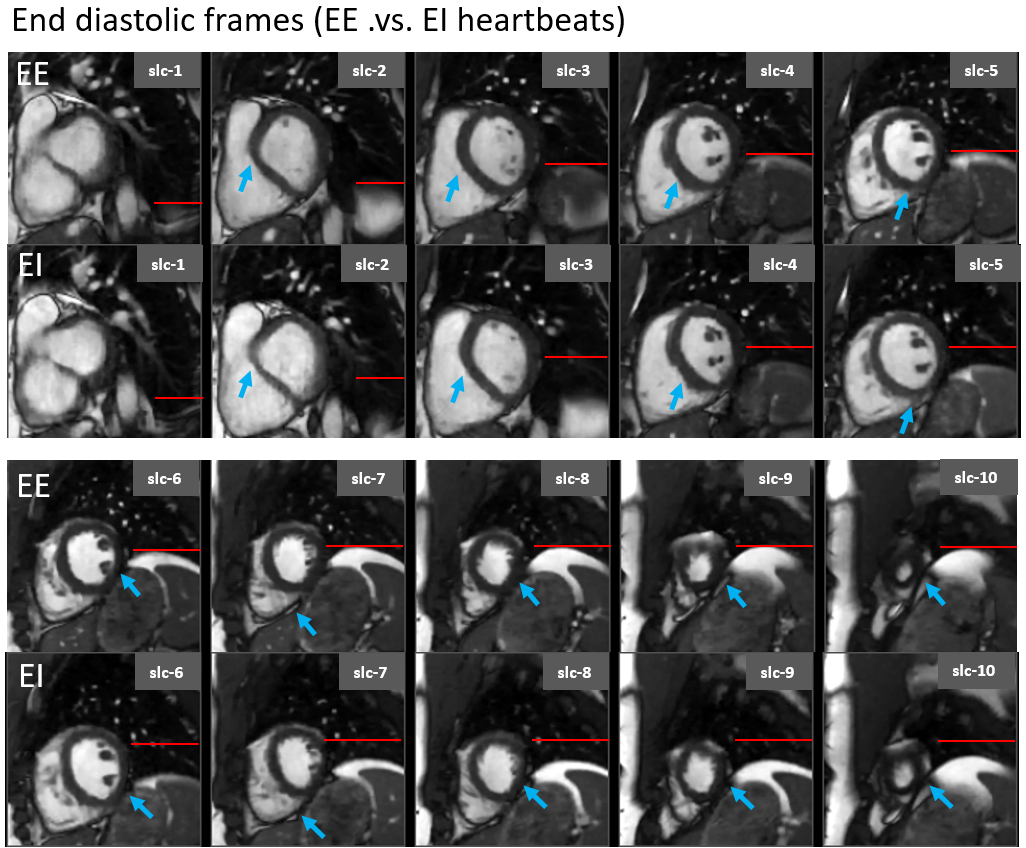}}
\caption{End expiration (EE) and end inspiration (EI) heartbeats identified using the proposed method. Both in-plane and through-plane motions, highlighted by the red lines and blue arrows, respectively, can be noticed. Only an end-diastolic frame is shown. Here, slc represents slice.}
\label{fig:res_image}
\end{figure}

Using the extracted respiratory signal as a guide, we extracted one end-inspiration (EI) and one end-expiration (EE) heartbeat from each slice for one of the volunteers. Fig.~\ref{fig:res_image} shows data separated into EE and EI heartbeats. All slices within EE have, as expected, elevated liver dome compared to the corresponding EI slices. However, a significant in-plane and through-plane motion of the heart can be noticed between EE and EI, highlighting the perils of mixing slices from different respiratory phases.


\section{Conclusion and Discussion}
\label{sec:typestyle}
We have proposed a new method to correctly assign the sign to the respiratory motion signal. The method utilizes the similarity of neighboring eigen images to adjust the sign of all slices with respect to the first slice. Then, the center-of-mass curve is used to adjust the global sign. The preliminary results are promising, where the proposed method corrected the sign successfully for all slices in all volunteers. \textr{Compared with the image registration method that is only effective with the properly selected ROI, the proposed} method does not require any manual intervention and is computationally fast. A potential issue for this study is related to the confinement of the respiratory signal to the second singular vector. As it has been reported previously~\cite{novillo2019unsupervised}, the presence of another motion source, e.g., the bulk motion from the subject, can threaten the localization of the respiratory motion to one singular vector. Another limitation of this study is the lack of validation in subjects with arrhythmia or inconsistent breathing. \textr{Since the cardiac motion is removed by the low-pass filter, we do not expect arrhythmias to impact the accuracy of this method. In the future, we will evaluate the proposed method in a larger population that includes subjects with inconsistent breathing patterns. We will also apply this method to other imaging orientations, e.g., long axis.}



\bibliographystyle{IEEEbib}
\bibliography{root.bib}

\end{document}